\pdfoutput=1
\documentclass[sigconf]{acmart}
\usepackage{bookmark}
\AtBeginDocument{%
  }

\usepackage[most]{tcolorbox}
\tcbuselibrary{skins, breakable, theorems}
\usepackage{xcolor}          
\usepackage{multirow}

\definecolor{myborder}{gray}{0.6}
\definecolor{mainbg}{gray}{0.97}
\definecolor{titlebg}{gray}{0.85}

\newtcolorbox{myrqbox}[1]{
    colback=mainbg,
    colframe=myborder,
    boxrule=1.5pt,
    arc=8pt,
    title=#1,
    fonttitle=\bfseries\sffamily,
    fontupper=\sffamily
}

\colorlet{myborder2}{blue!40!white}  
\colorlet{mainbg2}{blue!10} 
\colorlet{titlebg2}{blue!30} 

\newtcolorbox{myfindingbox}[1]{
    enhanced,
    colback=mainbg2,
    colframe=myborder2,
    boxrule=1.5pt,
    arc=8pt,
    attach boxed title to top left={
        xshift=1cm,
        yshift*=-\tcboxedtitleheight/2,
        yshiftadd=\tcboxrule/2,
    },
    title=#1,
    fonttitle=\bfseries\sffamily,
    fontupper=\sffamily,
    boxed title style={
        colback=titlebg2,
        colframe=myborder2,
        boxrule=1.5pt,
        rounded corners=4pt,
    }
}

\begin{document}

\title{One Size Fits All? An Empirical Comparison of ADR Templates regarding Comprehension, Usability, and Ease of Adoption}



\author{Fernando Nogueira}
\affiliation{%
  \institution{Federal University of Amazonas}
  \city{Manaus}
  \country{Brazil}}
\email{fernando.neves@icomp.ufam.edu.br}

\author{Nabson Silva}
\affiliation{%
  \institution{Federal University of Amazonas}
  \city{Manaus}
  \country{Brazil}}
\email{nabson.paiva@icomp.ufam.edu.br}

\author{Tayana Conte}
\affiliation{%
  \institution{Federal University of Amazonas}
  \city{Manaus}
  \country{Brazil}}
\email{tayana@icomp.ufam.edu.br}

\begin{abstract}
  \textbf{Context}: Documenting Architectural Design Decisions (ADDs) is a critical factor in the software lifecycle, essential for efficient system maintenance, developer onboarding, and preventing knowledge vaporization. Although various templates for Architectural Decision Records (ADRs) have been proposed, there is a lack of empirical evidence comparing them. \textbf{Goal}: To address this gap, this paper aims to identify which ADR template best supports comprehension, usability, and ease of adoption: Tyree/Akerman's template, Nygard’s ADR, arc42, Y-statements, and MADR. \textbf{Method}: We compared these templates using the DESMET FA method in a two-step evaluation. First, the two primary authors evaluated the five templates through the DESMET FA, based on their software architecture expertise. The two top-performing templates were then used as treatments in a controlled experiment conducted with undergraduate students. \textbf{Results}: In the feature analysis by experts, the top-performing templates were those of Nygard and MADR. In the subsequent controlled experiment, Nygard's template outperformed MADR in terms of the Overall Score. Qualitative analysis of participant feedback revealed the factors influencing template preference. The findings indicate that Nygard supports concise and objective documentation, while MADR facilitates structural details and specific architectural requirements. \textbf{Conclusion}: This paper provides an evidence-based strategy for ADR template adoption by offering a comparison between them. The findings present a decision-making guide that assists practitioners and researchers in selecting ADR templates aligned with project constraints, aiming to minimize documentation overhead and increase architectural knowledge retention. 
  
\end{abstract}



\maketitle

\section{Introduction}
Software architecture plays a pivotal role throughout the software development lifecycle~\cite{clements2003documenting, garlan2000software, kruchten2012strategic}, directly affecting maintainability and contributing to the accumulation of technical debt~\cite{besker2018managing}. These effects are particularly pronounced during the long-term evolution of software systems, where architectural decisions shape both system adaptability and sustainability~\cite{besker2017impact, DBLP:conf/ease/StolAB10, kruchten2004ontology}. Despite this relevance, software development practices have often emphasized architectural design activities while paying comparatively less attention to the systematic validation and documentation of architectural decisions~\cite{garlan2000software}. This imbalance has tangible consequences: rigorous architecture documentation supports effective maintenance and long-term knowledge preservation~\cite{ahmeti2024architecture, wan2023software}, whereas insufficient documentation hampers developer onboarding and intensifies the challenges associated with evolving complex software systems~\cite{nassif2025evaluating}.

The documentation of Architectural Design Decisions (ADDs)~\cite{capilla201610, clements2003documenting} plays a central role in preserving architectural knowledge over time. Architectural decisions are typically made by weighing the trade-offs among available technical alternatives~\cite{habli2007capturing}, and documenting this rationale enables development teams to later understand why a system was designed in a particular way. Beyond supporting retrospective understanding, ADD documentation informs future architectural decision-making~\cite{babar2006assessing}, helping teams avoid repeating past mistakes and mitigating the loss of architectural knowledge—often referred to as knowledge vaporization~\cite{jansen2005software, carvalho2025software, borowa2023rationales}.

Given the relevance of ADD documentation, several approaches and tools have been proposed and adopted in industrial settings~\cite{jansen2007tool, tyree2005architecture} to support the management of architectural decision knowledge. However, empirical evidence comparing these approaches remains limited, particularly regarding their adoption, comprehensibility, and usability. According to~\citet{jansen2005software}, these characteristics constitute essential quality attributes of effective decision documentation and are critical to its sustained use in practice.

The concept of Architectural Decision Records (ADRs) has gained traction in agile environments, leading to the proposal of multiple templates such as Tyree/Akerman~\cite{tyree2005architecture}, Nygard’s ADR~\cite{nygard2011adr}, and MADR~\cite{kopp2018markdown}. Despite their increasing adoption, the selection of an ADR template in industrial practice is often guided by ad hoc preferences, personal experience, or community conventions rather than systematic empirical evidence. This lack of evidence-based guidance hinders informed decision-making and increases the risk of miscommunication of architectural decisions to developers or stakeholders. Although prior work highlights the importance of comprehensibility, usability, and ease of adoption for effective decision documentation~\cite{jansen2005software}, empirical comparisons of ADR templates along these dimensions remain scarce.

This work compares ADR templates by applying the DESMET Feature Analysis (FA) method~\cite{kitchenham1996desmet} in a two-step evaluation. First, we applied the DESMET FA to screen five prominent templates, namely Tyree/Akerman~\cite{tyree2005architecture}, Nygard’s ADR~\cite{nygard2011adr}, arc42~\cite{arc42_template}, Y-statements~\cite{6576117}, and MADR~\cite{kopp2018markdown}, to identify those most suitable for agile contexts. We selected these templates based on their prevalence in the industry, as suggested by~\citet{zimmermann2015architectural}. This feature analysis resulted in the selection of Nygard and MADR as the top-performing templates. Subsequently, we conducted a controlled experiment with 33 undergraduate software engineering students as participants to compare these two templates. At this stage, we utilized the DESMET FA to calculate an Overall Score based on empirical metrics that cover comprehension, usability, and ease of adoption. By analyzing these scores, we assess which structural aspects contribute to a more effective balance between documentation effort and consumption, providing evidence-based guidance to practitioners to minimize documentation overhead. Furthermore, we performed a qualitative analysis of participant feedback to uncover the factors that influence template preference.


The contributions of this study assist practitioners and researchers in navigating the trade-offs of architectural documentation, providing:
\begin{itemize}
    \item An empirical comparison of widely used ADR templates through a standardized evaluation
    \item Evidence-based insights into how specific ADR structures impact documentation overhead and practitioner adoption
    \item Evidence-based recommendation to assist architects in the ADR adoption in agile environments
\end{itemize}

\section{Background}
This section explores the fundamental principles of \textbf{Architectural Knowledge (AK)}, \textbf{Architectural Decisions (ADs)}, and \textbf{Architectural Decision Records (ADRs)}. It establishes the theoretical foundation for understanding the context and significance of the proposed work.

\subsection{Architectural Knowledge}
Architectural Knowledge (AK) constitutes the comprehensive representation of the design logic, extending beyond static artifacts to include the design decisions, assumptions, context, and rationale driving the system's evolution~\cite{kruchten2006building}. Traditional documentation focuses primarily on the resulting components and connectors, often neglecting the reasoning behind their selection~\cite{perry1992foundations}. This omission creates a gap where critical knowledge remains implicit, residing solely in the architects' minds. The loss of this tacit information leads to \textit{knowledge vaporization}, a concept first described by \citet{jansen2005software} as the loss of embedded architectural knowledge. That highlights the problems arising from the loss of AK, such as violations of design rules and constraints, the maintenance of obsolete design decisions, and limitations on the reusability of architectural artifacts.

\subsection{Architectural Decisions}
Architectural Decisions (ADs) represent the core units of change in software architecture. \citet{kruchten2004ontology} propose treating ADs as first-class entities, shifting the architectural paradigm from a component-centric view~\cite{kruchten20024+,clements2003documenting} to a decision-centric one. An AD is not an isolated event but a structured entity possessing distinct attributes, including rationale, scope, state, cost, and associated risks. The ontology of ADs categorizes these decisions into distinct classes:
\begin{itemize}
    \item \textbf{Existence decisions:} Choices regarding the inclusion of structural elements or artifacts in the system.
    \item \textbf{Ban decisions:} Explicit exclusions of specific technologies or patterns to prevent technical debt or incompatibility.
    \item \textbf{Property decisions:} Establish design rules or guidelines that pervade the system, often addressing quality attributes.
    \item \textbf{Executive decisions:} Decisions related to the process, resources, or business constraints affecting the architecture.
\end{itemize}

ADs are often intertwined with each other~\cite{jansen2005software}. A decision may \textit{constrain}, \textit{enable}, \textit{conflict with}, or \textit{override} other decisions~\cite{kruchten2004ontology}. Capturing these relationships and the traceability to external artifacts is essential for maintaining the integrity of the system's design graph over time.

\subsection{Architectural Decision Records}
\begin{figure}[h]
  \centering
  \includegraphics[width=\linewidth]{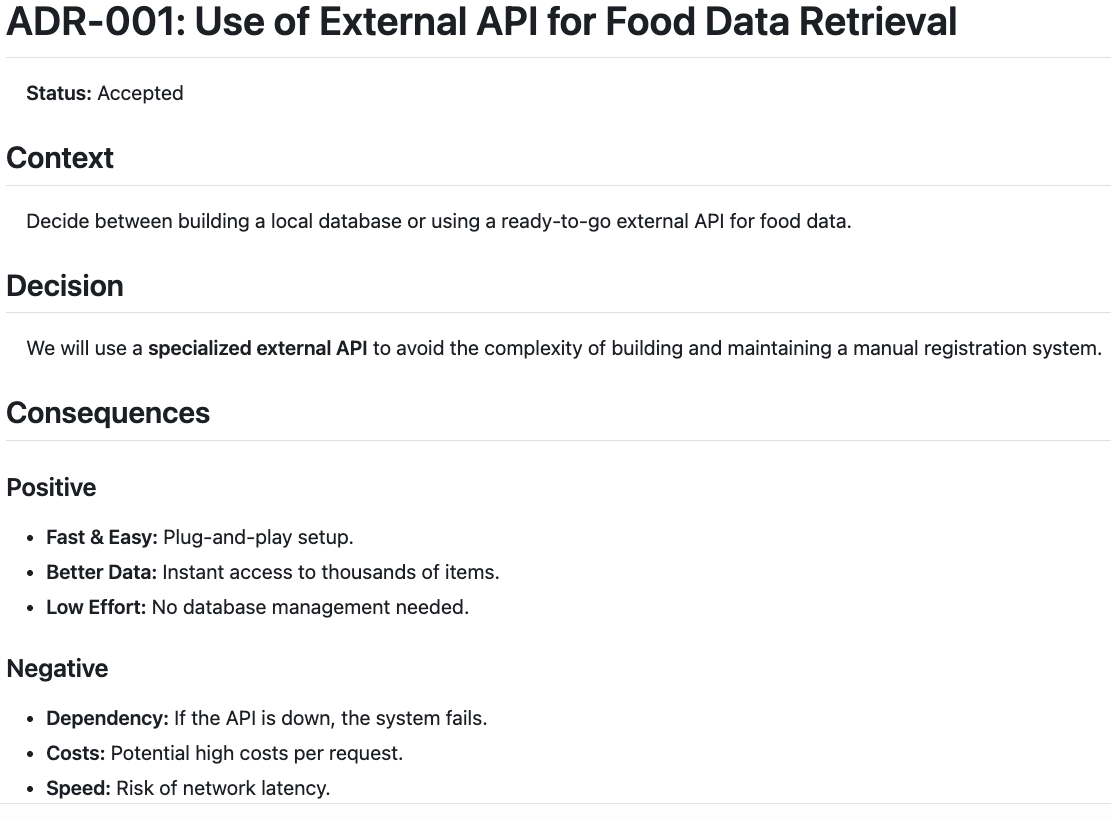}
  \caption{ADR example of a generic architectural decision using Nygard`s template.}
  \label{adr-example}

\end{figure}

Architectural Decision Records (ADRs) serve as the operational mechanism to externalize these complex AD entities. While ADs represent the ontological concept, ADRs provide the tangible format to persist them, facilitating the capture of rationale within a version-controlled workflow~\cite{nygard2011adr}. Although structured approaches were previously defined, such as the Architecture Decision Description by \citet{tyree2005architecture}, the specific concept of ``ADR'' emphasizing lightweight, text-based documents was popularized by \citet{nygard2011adr}. Figure \ref{adr-example} illustrates an instantiated decision record using Nygard's template, documenting the integration of an external API.

\subsection{Related Works}

The landscape of Architectural Decision Records (ADRs) has been studied from different perspectives, ranging from industrial surveys to the proposal of specific documentation templates.

\citet{ding2014open} conducted an exploratory survey of 2000 Open Source Software (OSS) projects to investigate how software architecture is documented in distributed and agile environments. The study analyzed the prevalence of architectural documents, the languages used, and the types of information captured (e.g., templates, rationale, and viewpoints). The study revealed that architectural rationale is rarely recorded due to the tension between documentation overhead and agile development, suggesting that developers prioritize simple, easy-to-maintain structures to facilitate coordination. However, the study does not provide any findings to overcome the overhead of documentation or to suggest ways to incorporate ADR adoption in the agile environment. Recent evidence by \citet{carvalho2025software} suggests a shift in this behavior of architectural documentation adoption. They conducted semi-structured interviews focused on the Brazilian software industry. The study revealed that senior practitioners actively document and review architectural decisions to support evolutionary design. Despite this, the study reveals a lack of systematic approaches or specific tools, relying instead on ad hoc documentation.

\citet{shahin2009architectural} provides one of the early systematic analyses of ADD models, identifying a consensus on core elements such as Decision, Rationale, and Solution, while highlighting variability in secondary elements across models. Their study primarily adopts a conceptual and comparative perspective, aiming to structure the ADD design space rather than to evaluate its practical use. As the work predates the widespread adoption of structured ADR templates, it does not address issues related to tool support, standardization of documentation practices, or empirical evidence of adoption. The study offers limited insight into how such models are used or perceived by practitioners.

\citet{ahmeti2024architecture} conducted an industrial case study with practitioners in a microservice-based environment to investigate the challenges of documenting architectural design decisions (ADDs) and the impact of introducing Architectural Decision Records (ADRs). The study followed an action research approach, identifying seven core challenges categorized into documentation culture, distributed system parts, knowledge transfer, and prioritization. To assess the transition, the authors compared pre- and post ``ADR intervention'' survey data and conducted interviews with developers and architects to map their observations to the identified challenges. As their results, the introduction of ADRs significantly improved team satisfaction and facilitated knowledge transfer by reducing reliance on individual memory. However, the study does not specify the ADR template used, although it acknowledges that using a different template would produce different results.

To the best of our knowledge, no papers specifically compare ADR templates, and only a few report on the structural effectiveness of these models. \citet{kopp2018markdown} introduces the MADR template, emphasizing its benefits for integrating a Markdown-based tool and achieving structural completeness. \citet{nygard2011adr} proposed the lean ADR template, which focuses on simplicity and agility and adapts decision architecture documentation to current industrial practices. \citet{zimmermann2015architectural} offers an impactful categorization of architectural knowledge management approaches, suggesting that the choice of a template should align with the project's needs, but does not provide comparative metrics for these approaches. And do not provide fundamental factors that drive practitioners to prefer one template over another. In general, existing research emphasizes the benefits of ADR adoption but lacks a focus on evaluating and comparing the attributes of ADR templates available in the literature.

\section{Study Design}
\begin{figure}[h]
  \centering
  \includegraphics[width=\linewidth]{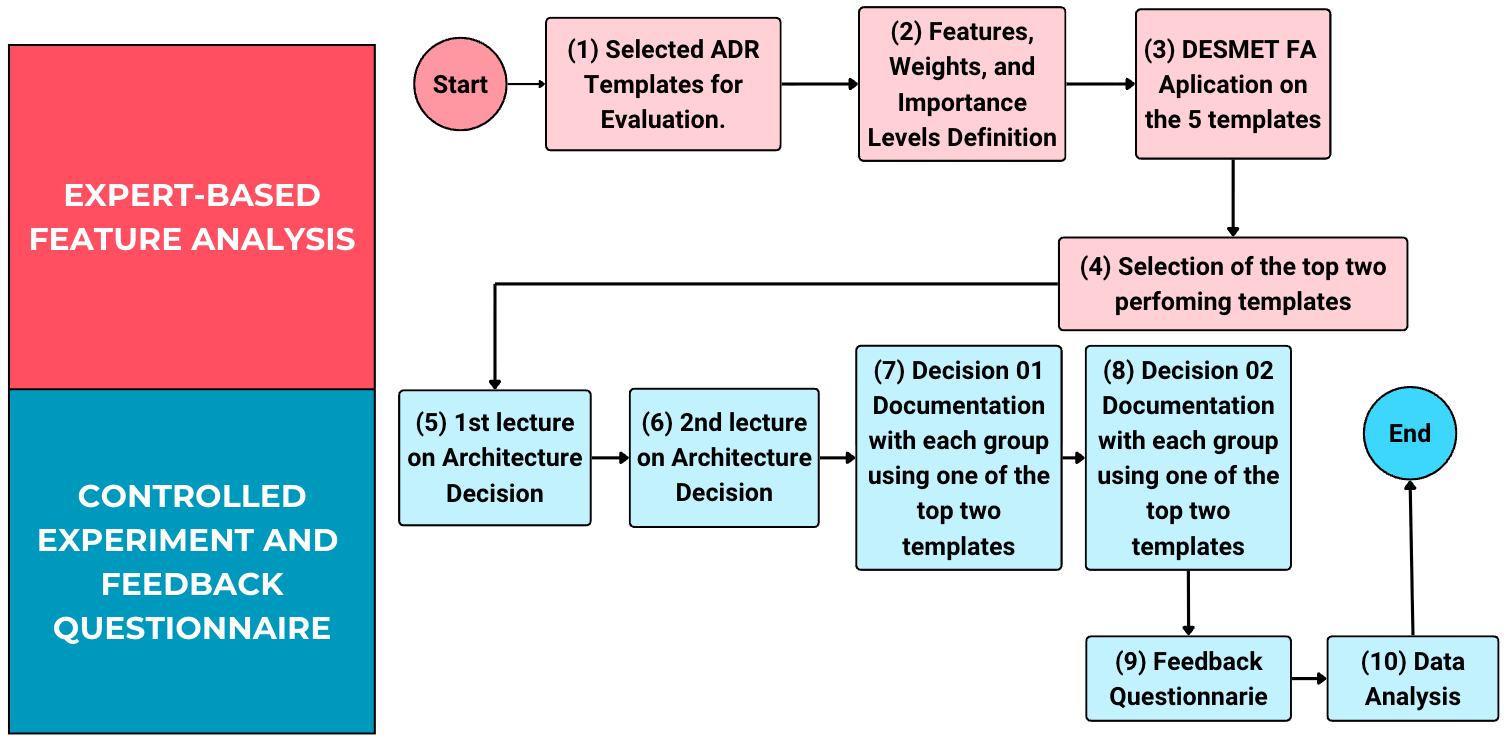}
  \caption{Two-step research design}
  \label{methodology}

\end{figure}

To achieve the study goal, we adopted a research design composed of an expert-based feature analysis followed by a controlled experiment. Initially, five ADR templates were assessed using the DESMET FA~\cite{kitchenham1996desmet} method to identify the most suitable templates. The two highest-ranked templates were then subjected to a controlled experiment with Software Engineering students. Figure \ref{methodology} illustrates the two-step research design.

\subsection{Goal and Research Questions}
\label{rqs}
The goal of this study is to evaluate and compare ADR templates for their ability to effectively support documentation and understanding of architectural decisions. The evaluation focuses on key quality attributes that influence the practical use of ADRs, namely comprehension, usability, and ease of adoption. Based on this, we formulate the following Research Questions (RQs):

\begin{myrqbox}{RQ1}
Which of the selected ADR templates is better according to comprehension, usability, and ease of adoption?
\end{myrqbox}

To address RQ1, we employed the DESMET FA in a two-step evaluation process based on three dimensions: comprehension, usability, and adoption. First, the two primary authors conducted a feature analysis of five candidate ADR templates to identify the top performers based on expert evaluation (detailed in Section \ref{preliminary-screening}). In the second stage, we conducted a controlled experiment with 33 software engineering students to validate the two selected templates, Nygard and MADR, as described in Section \ref{planning-ce}. This stage utilized the same DESMET criteria to aggregate the experimental results into a quantitative Overall Score

\begin{myrqbox}{RQ2}
What are the key factors that influence students' preferences when choosing an ADR template?
\end{myrqbox}
To address $RQ_2$, we collected qualitative feedback from the participants of the controlled experiment to identify the factors that influence their preference for one ADR template over another. These data were analyzed following the Grounded Theory procedures proposed by \citet{corbin2014basics}, specifically focusing on open and axial coding.

\subsection{Expert-based Feature Analysis}
\label{preliminary-screening}
Before conducting the controlled experiment, we performed the DESMET FA to evaluate and select the most suitable ADR templates from a set of candidates. This evaluation was conducted by the first two authors, who are domain experts with over four years of industrial experience in software architecture. DESMET is a comprehensive methodology for evaluating software engineering methods and tools proposed by \citet{kitchenham1996desmet}. We chose the Feature Analysis modality because it allows systematic comparison of multiple artifacts based on their characteristics, especially when resources for large-scale experimentation across all candidates are limited.

We employed DESMET FA to evaluate the initial set of five ADR templates and select the two top-performing templates for further testing with students. This step involved: (i) defining a set of relevant features for architectural documentation; (ii) assigning weights and importance levels to each feature based on architectural expertise; and (iii) scoring each candidate template using judgment scales. The results of this evaluation ensured that the controlled experiment focused on templates with the highest potential for effective documentation in software projects, as determined by the overall score.

\subsubsection{Selected ADR Templates for Evaluation}
The ADR templates selected for this study are widely used in industry and were chosen based on their public accessibility and documented use in professional projects. Following the rationale of~\citet{zimmermann2015architectural}, our selection does not aim to be complete but rather representative of current industry practices. Therefore, purely academic or scientific approaches were excluded. An additional selection criterion was the popularity of these templates within the open-source community, as evidenced by their inclusion in the widely recognized ADR  curation repository\footnote{\url{https://github.com/joelparkerhenderson/architecture-decision-record}}. Thus, the following templates were selected for evaluation: the Tyree and Akerman template~\cite{tyree2005architecture}, M. Nygard's ADR~\cite{nygard2011adr}, the arc42 resource by Hruschka and Starke~\cite{arc42_template}, the Y-statements by Zimmermann et al.~\cite{6576117}, and the Markdown ADR (MADR) by Kopp et al.~\cite{kopp2018markdown}. Table~\ref{tab:adr-template-comp} summarizes the main structural characteristics and their key focus of the five selected templates.

\begin{table*}
\caption{Comparison of the selected ADR templates for evaluation}
\label{tab:adr-template-comp}
\begin{tabular}{@{}lllll@{}}
\toprule
ADR Template                                                    & \multicolumn{1}{l}{Publication year} & Expected Length                             & Format      & Key Structural Focus                 \\ \midrule
Tyree/Akerman~\cite{tyree2005architecture} & 2005                                 & {\color[HTML]{1F1F1F} 1–2 pages}            & Simple text & Detailed rationale and implications. \\
Nygard ADR~\cite{nygard2011adr}           & 2011                                 & {\color[HTML]{1F1F1F} 3–5 short paragraphs} & Simple text & Minimalist, log-based versioning.    \\
arc42~\cite{arc42_template}             & 2012                                 & {\color[HTML]{1F1F1F} Multi-section (Long)} & Multi-form  & Full architectural integration.      \\
Y-Statements~\cite{6576117}                & 2013                                 & {\color[HTML]{1F1F1F} Single Sentence}      & Simple Text & High-level decision summary.         \\
MADR~\cite{kopp2018markdown}           & 2018                                 & {\color[HTML]{1F1F1F} 1 page (Structured)}  & Markdown    & Options comparison and pros/cons.    \\ \bottomrule
\end{tabular}
\end{table*}

\subsubsection{Features, Weights, and Importance Levels Definition}
\label{fs-and-metrics}
Following the DESMET FA methodology~\cite{kitchenham1996desmet}, we identified and categorized the required features (comprehension, usability, and ease of adoption) into three Feature Sets (FS). Each set was assigned an Importance Weight ($IW$) through an iterative consensus process among the researchers. Table \ref{tab:fa} summarizes the features, weights, and importance levels of the Sub-features of each FS. 

To ensure comparability between different measurement units (e.g., time, Likert scales), all metrics were normalized to a common scale ($0$ to $1$) and synthesized into an Overall Score. The FS and their specific measurement protocols are defined as follows:

\begin{itemize}
    \item \textbf{FS1: Structural Comprehension}: This set evaluates the participant's \textit{comprehension} of the template. Decision, Context, and Consequences are essential fields of an ADR. Therefore, we assume that if participants can easily identify and map these essential elements, they have successfully comprehended the template and its structure. We measured this based on the identifiability of key fields on a 3-point ordinal scale: 1 (``Identified''), 0.5 (``Partially Identified''), and 0 (``Not Identified''). Although fundamental, this set received the lowest weight because most industry-standard templates already satisfy these requirements~\cite{zimmermann2015architectural}.

    \item \textbf{FS2: Efficiency and Usability}: This set assesses \textit{usability} by evaluating how easy and objective it is for the participant to document a decision using the template. We assume that a usable template reduces cognitive load, allowing users to complete the task faster and with less reliance on external help. Therefore, we aggregate two key metrics: (i) \textit{Efficiency}, quantified by the documentation completion time and the frequency of consulting supporting materials (both inverted during normalization so that lower time and fewer consultations contribute positively to the score), and (ii) \textit{Perceived Objectivity}, measured on a 5-point Likert scale ranging from ``Least Objective'' (1) to ``Most Objective'' (5). This set received the highest Importance Level ($IL=4$) because reducing ``documentation overhead'' is critical for agile adoption~\cite{zimmermann2015architectural, kopp2018markdown, yang2019integrating}.
    
    \item \textbf{FS3: Adoption Potential}: To understand whether participants found the template \textit{easy to adopt}, we asked them if they would recommend the adoption of the template. We measured this using a scale in the $[0,10]$ interval. According to ~\citet{kitchenham1996desmet}, user preference is a decisive indicator of a method's desirability. Therefore, this set received a high Importance Level of $IL=4$.
\end{itemize}

By applying the DESMET FA scoring rubric to these weighted features, we calculated each model's overall performance using the equations \ref{sfs-equation} and \ref{os-equation}. This quantitative method served as the basis for both the initial expert feature analysis and the subsequent statistical analysis of the controlled experiment. 

\begin{equation}
\label{sfs-equation}
\text{$Score$}_{FS_x} = \frac{\sum (\text{$IL$}_{FS_x} \times \text{$SFJ$}_x)}{\text{$PM$}_{FS_x}}
\end{equation}

\begin{equation}
\label{os-equation}
\text{$OS$} = \sum (\text{$Score$}_{FS_x} \times \text{$IWF$}_{FS_x})
\end{equation}

   $\text{$IL$}_{FS_x}$: Importance Level of FS
   
   $\text{$SFJ$}_x$: Sub-feature judgement (value of the sub-feature metric)
   
   $\text{$PM$}_{FS_x}$: Sum of the importance levels of the sub-features
   
  $\text{$Score$}_{FS_x}$: FS score
  
   $\text{$IWF$}_{FS_x}$: FS importance weight
   
   $\text{$OS$}$: Overall Score

\begin{table*}
\caption{Feature Sets and subfeatures evaluated in the FA.}
\label{tab:fa}
\begin{tabular}{@{}lllll@{}}
\toprule
ID  & Feature set              & Sub-feature                                                                                                                                 & Importance Level (IL) & Importance Weight (IW) \\ \midrule
FS1 & Structural Comprehension  & \begin{tabular}[c]{@{}l@{}}FS1.01 - Presence of a dedicated section\\  to describe the decision\end{tabular}                                & Interesting (1)       & \multirow{3}{*}{0.15}  \\
    &                          & \begin{tabular}[c]{@{}l@{}}FS1.02 - Presence of a dedicated section\\ to describe the context of the decision\end{tabular}                  & Interesting (1)       &                        \\
    &                          & \begin{tabular}[c]{@{}l@{}}FS1.03 - Presence of a dedicated section \\ to describe the consequences/outcomes\\ of the decision\end{tabular} & Interesting (1)       &                        \\ \midrule
FS2 & Efficiency and Usability & \begin{tabular}[c]{@{}l@{}}FS2.01 - Total time required to complete\\ the documentation task.\end{tabular}                                  & Mandatory (4)         & \multirow{3}{*}{0.50}  \\
    & \textbf{}                & \begin{tabular}[c]{@{}l@{}}FS2.02 - Frequency of external consultations\\ during the task.\end{tabular}                                     & Mandatory (4)         &                        \\
    & \textbf{}                & FS2.03 - Degree of the perceived objectivity                                                                                                & Mandatory (4)         &                        \\ \midrule
FS3 & Adoption Potential       & \begin{tabular}[c]{@{}l@{}}FS3.01 - Likelihood of recommending\\  the model.\end{tabular}                                                   & Mandatory (4)         & 0.35                   \\ \bottomrule
\end{tabular}
\end{table*}

\subsection{Controlled Experiment Planning}
\label{planning-ce}

We planned the controlled experiment using the top-two performing ADR templates resulting of the previous step.

\subsubsection{Context Selection}

We conducted the experiment with undergraduate students in software engineering enrolled in a course on Software Architecture and Architectural Decisions. This context simulates junior developers entering a company, where they are required to work with and document Architectural Decisions using the selected templates. The practical session took place after students had received theoretical instruction on the AD documentation process.

\subsubsection{Variable Selection} The independent variable is the ADR Model Template, with two treatments: (1) the Nygard template and (2) the MADR template.
The dependent variable is the Overall Score of the DESMET FA, derived from metrics collected from students to evaluate three main constructs. To measurements, we applied the same metrics defined in the section \ref{fs-and-metrics}.
That enables us to statistically compare them~\cite{wohlin2012experimentation}.

\subsubsection{Selection of subjects}  Participant selection was based on convenience sampling~\cite{wohlin2012experimentation}, targeting undergraduate Software Engineering students enrolled in the Software Architecture course. The course provided the necessary background for the study, covering software modeling (UML use case, class, and sequence diagrams), architectural styles (e.g., Layered, MVC, and Microservices), and design patterns. Participation in the study was voluntary, and only students who signed the consent form and completed all required experimental tasks were included.

\subsubsection{Hypothesis Formulation} We formulated one primary null hypothesis ($H_0$) and its corresponding alternative hypothesis ($H_1$) to cover the RQ1.
\begin{itemize}
\item \textbf{$H_0$ (Null Hypothesis)}: There is no statistically significant difference in the DESMET FA Overall Scores between Nygard’s model and the MADR model.
\item \textbf{$H_1$ (Alternative Hypothesis)}: There is a statistically significant difference in the DESMET FA Overall Scores between Nygard’s model and the MADR model.
\end{itemize}

\subsubsection{Experiment design} 
\begin{figure}[h]
  \centering
  \includegraphics[width=\linewidth]{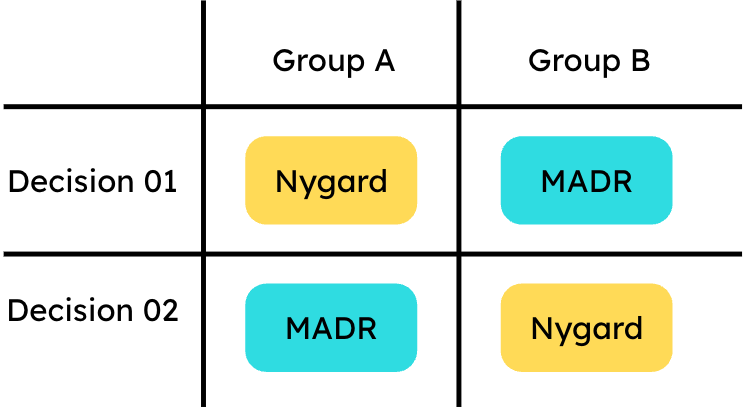}
  \caption{Experiment Design Illustration.}
  \label{design-exp}
\end{figure}

Students completed two documentation tasks. We designed two architectural decision scenarios (Objects) to support the tasks: Decision 1, related to API communications, and Decision 2, related to the database type. Students were familiar with both decisions because they were related to exercises they had done in previous classes to ensure the decision would not affect the templates comparison. To compare the proposed ADR templates (Treatments), the Nygard template, and the MADR template, we employed a crossover design~\cite{vegas2015crossover}. The students were divided into two groups, balanced on their experience as determined from the characterization feedback: Group A and Group B. During the first task (Day 1), Group A documented Decision 1 using the Nygard template, while Group B documented Decision 1 using the MADR template. To minimize fatigue effects, we introduced a one-week interval between the tasks. One week later, for the second task (Day 2), the treatments were crossed over: Group A documented Decision 2 using the MADR template, and Group B documented Decision 2 using the Nygard template, to systematically balance the order of application and mitigate learning effects~\cite{wohlin2012experimentation}. Figure~\ref{design-exp} illustrates the experiment design.

\subsubsection{Instrumentation} 
\begin{figure}[h]
  \centering
  \includegraphics[width=\linewidth]{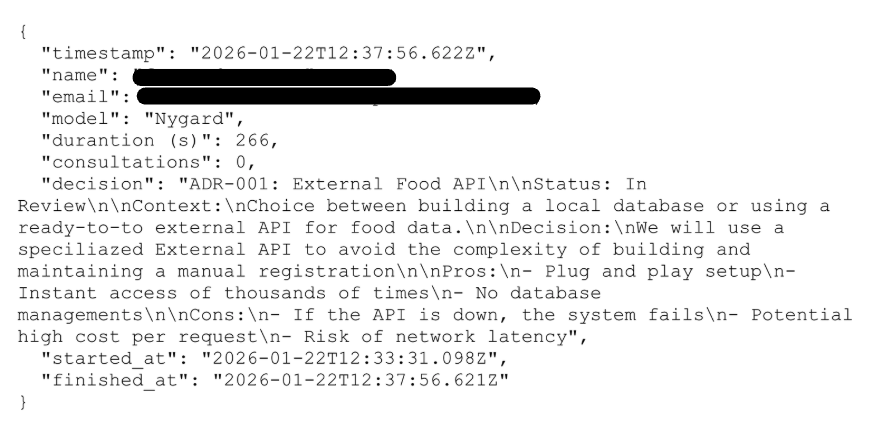}
  \caption{Example of Report Log generated by the platform.}
  \label{rep-log}
\end{figure}

The instruments for this experiment include a set of forms, training materials, architectural decision scenarios, a software platform, and specific task guidelines. We describe each instrument as follows:

\begin{itemize}
    \item \textbf{Training}: A presentation delivered during a class session explaining the core concepts of Architectural Decisions (ADs) and the AD documentation process, followed by a hands-on activity in which students document the design decisions identified within a generic e-commerce project.
    \item \textbf{Consent Form}: A form that outlined the purpose of the experiment and detailed how it would be conducted. We emphasized that participation was voluntary and explained that the collected data would be used for analysis and scientific publications. Participants signed the consent form before the experiment began.
    \item \textbf{Characterization Form}: A set of questions designed to gather information about the participants' profiles, including their working experience in software architecture and their familiarity with designing software systems, to ensure the balancing of the experimental groups.
    \item \textbf{Objects (Architectural Decision Scenarios)}: Descriptions of two standardized architectural decisions that students had to document using the provided templates. Both decisions were previously discussed in the class sessions.
    \begin{itemize}
        \item \textbf{Decision 01}: described a scenario about choosing how to communicate with an external API for a generic system.
        \item \textbf{Decision 02}: described a scenario about selecting the Postgres database for a generic system.
    \end{itemize}
    \item \textbf{ADR Documentation Platform and Guidelines}: We provided an online platform for students to write the ADRs. This platform also hosted the support materials and specific task instructions for both the Nygard template (Nygard's Guidelines) and the MADR template (MADR's Guidelines). The platform automatically captured the total time spent (Time to document) and the number of times the support material was accessed (Number of consultations). This data was saved in a report log file, returned after the documentation of the decision. Figure \ref{rep-log} illustrates an example of a report log generated.
    \item \textbf{Data Collection Form}: A post-task questionnaire used to collect data to calculate the dependent variables. This form included: (a) questions for field recognition (if the model had fields for decision, context, and consequences); (b) the Likert scale for perceived objectivity; and (c) the 0--10 recommendation score. This form also required participants to upload the log file generated by the documentation platform.
    \item \textbf{Feedback Form}: A post-experiment questionnaire was used to collect participants’ qualitative feedback on their experience using the ADR templates. The questionnaire included open-ended questions, aiming to assess their perceptions regarding ease of use, clarity of the template structure, and overall satisfaction with the documentation process. 

\end{itemize}

\subsection{Execution}

To ensure the realism and representativeness of the documentation tasks, we conducted a pilot study with one volunteer to review the selected scenarios and the instruments. The pilot study confirms that the description of the scenarios was clear and illustrative of common challenges in daily architecture. The main adjustments were made to the feedback questionnaire. For example, we added images on the questionnaire to help the participant describe their perception of the template's structure

\begin{figure}[h]
  \centering
  \includegraphics[width=\linewidth]{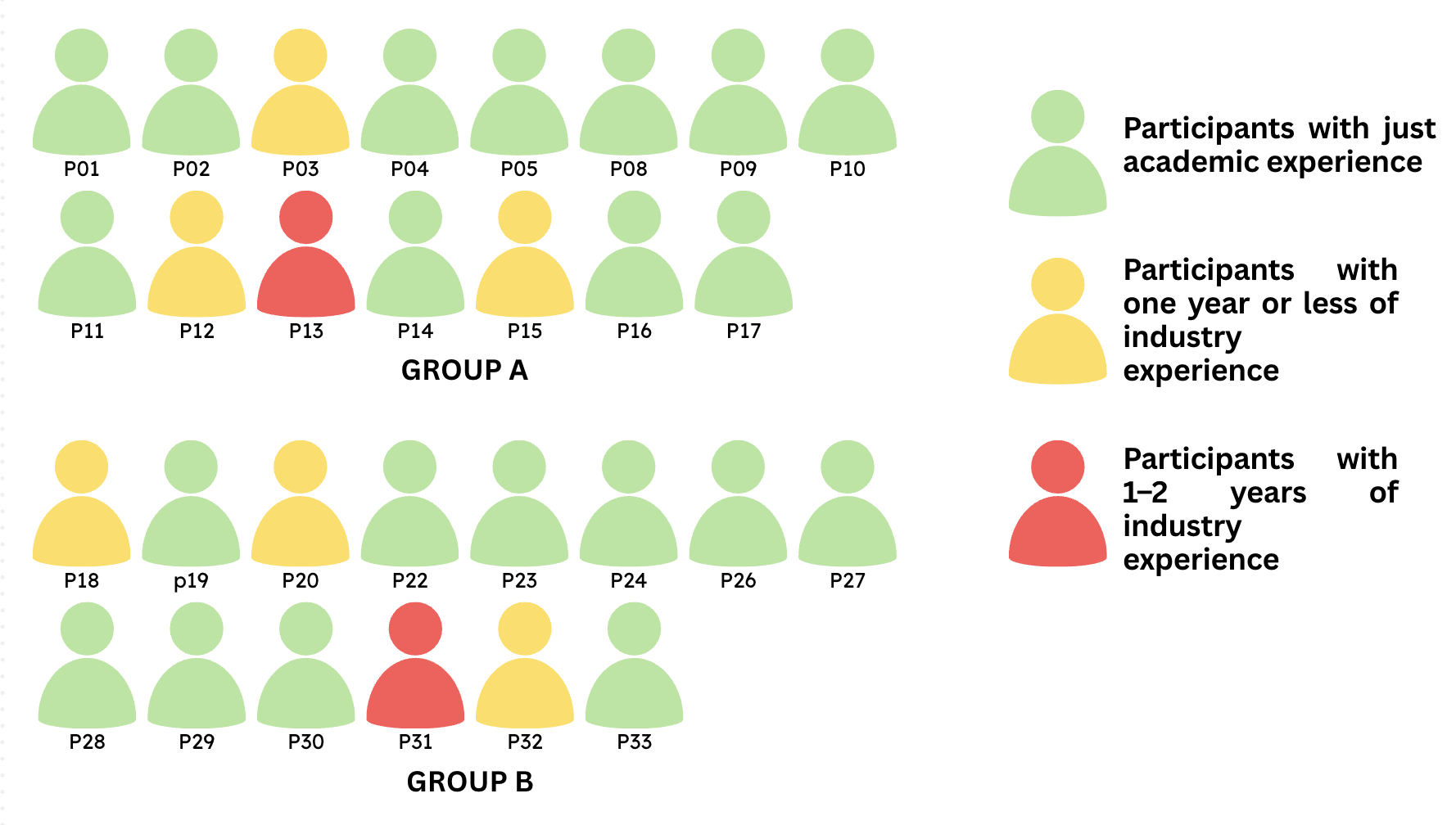}
  \caption{Participants distribution.}
  \label{balancing}
\end{figure}

Before the experiment, students attended two lectures covering software architecture patterns, templates, and decision-making.To ensure practical familiarity, participants also completed a hands-on documentation exercise during the final lecture.. Following the second lecture, students were invited to participate voluntarily by signing a consent form and completing a characterization questionnaire. We utilized the characterization data to balance the groups through block randomization, ensuring that participants' expertise levels and development experience were distributed approximately equally between the groups. A total of 33 students participated (17 in Group A and 16 in Group B). Figure \ref{balancing} illustrates the groups' balancing.

The experiment followed a crossover design, with two sessions of one and a half hours each, one week apart. The groups were also in two different laboratories, where we equally followed the experiment's guidelines. In the first session, both groups documented the first architectural decision: Group A utilized the Nygard template, while Group B used the MADR template. In the second session, the treatments were crossed: Group A used the MADR template and Group B used the Nygard template to document a second architectural decision. After completing both tasks, participants provided qualitative feedback via an exit questionnaire.

\subsection{Analysis and Interpretation} 
Before conducting the data analysis, we excluded responses from students who missed two prerequisite lectures on architectural decision documentation and from those who were absent on the first day of the experiment. This exclusion was necessary because their incomplete participation would introduce biases (such as learning effects) that the crossover design was intended to mitigate.

\subsubsection{Quantitative Analysis} Since each participant evaluated both the Nygard and MADR templates (paired samples), we performed a comparative analysis to test our primary null hypothesis ($H_0$). 

As the study used a crossover design, each participant evaluated both templates, yielding paired observations. Therefore, all analyses were conducted using paired statistical tests. We began with a visual inspection of the Overall Score distributions using boxplots to identify potential differences in central tendency and dispersion between the two templates.

To select the appropriate inferential test, we applied the Shapiro–-Wilk test to verify the normality of the paired samples separately, as recommended by Wohlin et al.~\cite{wohlin2012experimentation}. When the normality assumption was met, we employed the Paired t-test. Otherwise, the Wilcoxon Signed-Rank Test was used. These tests were applied to the Overall Scores to determine if the difference between Nygard and MADR was statistically significant ($p < 0.05$).

\subsubsection{Qualitative Analysis} To complement the quantitative findings and address $RQ_2$, we analyzed student feedback using the systematic procedures of Grounded Theory (GT)~\cite{corbin2014basics}. We first applied open coding, which involved creating codes representing relevant concepts regarding template usability and documentation support.  Then, when applying axial coding, we identified connections among the codes and grouped them into broader categories to reveal the underlying factors influencing template preference and the strengths/limitations of each model. Since our objective was to explore the factors influencing ADR selection rather than to generate a formal theory, we did not perform selective coding, providing a detailed thematic characterization of the results.

\section{Results}
This section presents the results for each RQ outlined in Section \ref{rqs}. For RQ1, we analyzed the quantitative data collected and calculated during the experiment, using statistical tests and boxplot visualizations to compare the samples. For RQ2, we summarized the insights from students' qualitative feedback.

\subsection{RQ1: Which of the selected ADR templates is better according to comprehension, usability, and ease of adoption?}
To address RQ1, we evaluated the DESMET FA Overall Scores obtained from the documentation tasks. This involved an expert-based feature analysis of all ADR templates conducted by the authors, after which the two templates with the highest DESMET FA Overall Scores were selected for use in the controlled experiment with students.

\subsubsection{Expert-based Feature Analysis} The Expert-based Feature Analysis results (Table \ref{tab:desmet-result-fa}) show that MADR and Nygard earned the highest overall scores, 0.900 and 0.868. This was largely due to their strong performance in Efficiency ($FS2$) and Adoption Potential ($FS3$). In contrast, while templates like arc42 and Tyree/Akerman are structurally complete ($FS1=0.15$), they incur significant efficiency penalties due to their high documentation overhead. To validate these findings, the first two authors tested the templates on two distinct architectural decisions. The results were consistent: MADR and Nygard ranked top in both cases, leading us to select them for the final controlled experiment.

\begin{table*}[]
\caption{Aggregated DESMET FA Results for Expert-based Feature Analysis (Mean Scores).}
\label{tab:desmet-result-fa}
\begin{tabular}{@{}lllll@{}}
\toprule
ADR Template  & OSFS1 (IW=0.15) & OSFS2 (IW=0.50) & OSFS3 (IW=0.35) & Overall Score \\ \midrule
MADR          & 0.15         & 0.450        & 0.300        & 0.900         \\
Nygard        & 0.15         & 0.441        & 0.277        & 0.868         \\
Y-statement   & 0.15         & 0.430        & 0.199        & 0.779         \\
Tyree/Akerman & 0.15         & 0.311        & 0.300        & 0.761         \\
arc42         & 0.15         & 0.101        & 0.119        & 0.370         \\ \bottomrule
\end{tabular}
\end{table*}

\subsubsection{Controlled Experiment} We began by verifying the distribution of the data to select the appropriate inferential statistical test.

We first analyzed the distribution of the Overall Score for both paired samples. Performing the Shapiro-Wilk Test on the first sample yields a $p-$value of  $0.0177$. For the second sample it yields a $p-$value of $0.0718$. Given a significance level of $\alpha = 0.05$, we rejected the null hypothesis of normality for the first sample and accepted for the second sample. Consequently, we selected the Wilcoxon Signed-Rank Test, a non-parametric test suitable for dependent samples with non-normal distributions, as recommended by Wohlin et al.~\cite{wohlin2012experimentation}.

The Wilcoxon signed-rank test revealed a statistically significant difference between the Nygard and MADR templates in terms of Overall Score ($W = 84.0$, $p = 0.002$), indicating that we can reject the null hypothesis $H_0$. To assess the practical significance of this result and address concerns about statistical power given the sample size ($n=33$), we calculated non-parametric effect sizes following~\citet{kitchenham2017robust}. We focused on Cliff's Delta ($\delta$) and the Probability of Superiority ($\hat{P}$), as these metrics do not assume normality and are robust to small sample sizes.

The estimated value of $\delta$ is $0.6364$ with a 95\% confidence interval of $[0.2842, 0.7011]$. According to established thresholds of ~\citet{kraemer2006size} and \citet{vargha2000critique}, this corresponds to a large effect size. This suggests that documentation produced with the Nygard template is more likely to achieve higher overall scores than documentation produced with the MADR template. This result is further supported by the probability of superiority ($\hat{P}$) of $0.8182$, indicating an $81\%$ probability that the Nygard template outperforms MADR in a random paired observation. This result indicates that the two templates differ in their overall performance across the quality attributes considered in RQ1, namely comprehension, usability, and ease of adoption.

\begin{figure}[h]
  \centering
  \includegraphics[width=\linewidth]{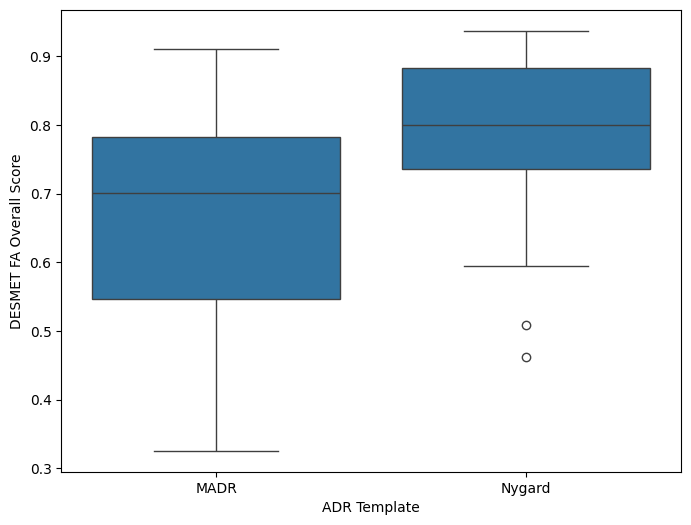}
  \caption{Boxplots presenting the data distribution of the overall score per template.}
  \label{fig:boxplot-overall}
\end{figure}

As illustrated in Figure~\ref{fig:boxplot-overall}, the Nygard template achieved a higher median Overall Score (approximately 0.8) compared to MADR (approximately 0.7). Since the Overall Score aggregates empirical measures related to the three quality dimensions, this result suggests that participants perceived the Nygard template as providing better overall support for understanding, using, and adopting architectural decision records than MADR.

\subsection{RQ2: What are the key factors that influence students' preferences when choosing an ADR template?} To address $RQ2$, we used ATLAS.ti\footnote{\url{https://atlasti.com/}} to analyze qualitative feedback through open (highlighted with bold text) and axial coding~\cite{corbin2014basics}. This process allowed us to pinpoint why participants preferred one ADR model over another. From the relationships between these codes, three primary categories that dictate the choice between Nygard and MADR emerged: Structural Granularity, Temporal Constraints, and Project Criticality.

\subsubsection{Structural Granularity}
The template's structural design was the most significant factor. Participants perceived a clear trade-off between the descriptive power of MADR and the directness of Nygard.

\textbf{MADR: Rationale Completeness and Ambiguity Reduction}. Many participants felt that \textit{MADR explains the decision better because of its greater number of fields, leading to less ambiguity}. As P13 noted: ``Due to the greater amount of information available regarding the architectural decision, MADR explains the reasons for the decision more clearly and effectively.'' Similarly, P18 suggested that a \textit{more robust structure forces a detailed description}: ``I believe that MADR, by including more topics, avoids ambiguity and also suggests a more detailed description of the decision to be made.'' This structure provides a better decision rationale, though P14 acknowledged it \textit{is more demanding}: ``The MADR, despite being longer and more demanding, manages to document the entire process more completely, leading up to a specific decision''.

\begin{myfindingbox}{}
\textbf{Finding 1}: MADR's structural granularity minimizes ambiguity and ensures rationale completeness, although it increases documentation effort and cognitive demand.
\end{myfindingbox}

\textbf{Nygard: Objectivity and Lean Documentation}. Conversely, Nygard was praised for being \textit{straight to the point}. Participants noted that \textit{Nygard is more objective because it has fewer fields, which prevents the document from feeling cluttered}, P32:``The amount of additional information makes the document feel more cluttered than the other model''. P12 illustrated this preference: ``I found Nygard to be more `straight to the point', in a positive way; it covers the important points and makes the decision very clear, and it's easier to fill out.'' P11 complemented this by stating: ``Nygard appears to be simpler and more objective in describing the architectural decision, making it easier to understand.''

\begin{myfindingbox}{}
\textbf{Finding 2}: Nygard’s straight-to-the-point structure promotes documentation objectivity and clarity, making it easy to understand the ADR template structure.
\end{myfindingbox}

\subsubsection{Temporal Constraints} 
The time available to document decisions emerged as another deciding factor. The results suggest that the ``documentation overhead'' it's one of the major concerns for the participants.

\textbf{Short time Scenarios}. \textit{For teams with short development cycles, Nygard's low overhead was seen as a decisive advantage}. P15 stated: ``I’d go with Nygard for daily tasks because of its low overhead; it’s faster to produce and consume, getting me the right info in less time'' P03 reinforced this perspective, noting that ``Nygard is faster and simpler to fill out'' which facilitates adoption in a short time to develop environments.

\textbf{Scenarios with more time available}: In contrast, participants felt that \textit{teams with more time available should invest in the MADR template}. P05 observed that ``MADR, on the other hand, has more optional fields and is better structured; while this makes it more time-consuming, it is also more objective''. This suggests that participants view the time invested in MADR as a trade-off for increased documentation quality and structure.

The time provided to develop a project was also stated as a definitive factor to choose a more adequate ADR template, as \textit{Teams with a short time to develop} would prefer Nygard's template, as stated P15: ``I’d go with Nygard for daily tasks because of its low overhead; it’s faster to produce and consume, getting me the right info in less time''. and emphasized by a quote from P03: ``Nygard is faster and simpler to fill out''. Whether \textit{Teams with more time available would prefer the MADR template}, as P05 noted: ``MADR, on the other hand, has more optional fields and is better structured; while this makes it more time-consuming, it is also more objective''. 

\begin{myfindingbox}{}
\textbf{Finding 3}: The selection of an ADR template is directly influenced by temporal constraints. Nygard minimizes overhead in short development cycle scenarios, whereas MADR is preferred when the project allows for a higher investment in structured documentation.
\end{myfindingbox}

\subsubsection{Project Scale} The perceived ``size'' of the business matter of the project also influenced the selection. The choice of an ADR template seems to scale with the Project Size and the Target Audience. 

\textbf{Greater-Scale Projects:} MADR was identified as the \textit{best option for larger projects or presentations to leadership}. P18 highlighted this distinction: ``However, for larger-scale projects or high-stakes presentations to leadership, I would opt for the MADR.''

\textbf{Small-Scale Projects:} For \textit{short projects, Nygard was deemed more suitable}. As P18 summarized: ``I personally liked both templates; however, for small-scale projects, I think Nygard would be my go-to''. This indicates that for less complex environments, the simplicity of the documentation outweighs the need for exhaustive fields

\begin{myfindingbox}{}
\textbf{Finding 4}: Project scale and audience profile determine the required level of documentation formality. MADR is preferred for large-scale projects, while Nygard is selected for small-scale projects to maintain agility.
\end{myfindingbox}

\section{Discussion}
The results of this study provide empirical evidence that the choice of an ADR template is not a one-size-fits-all decision, but rather depends on a set of contextual trade-offs involving structural granularity, temporal constraints, and project scale. These findings help explain why prior studies report both the benefits of ADR adoption and the lack of consensus on how architectural decisions should be documented in practice.

The preference patterns observed between MADR and Nygard highlight a fundamental trade-off between documentation completeness and cognitive overhead. Participants consistently perceived MADR’s higher structural granularity as beneficial for reducing ambiguity and ensuring rationale completeness, illustrated by Findings 1 and 2, corroborating earlier conceptual analyses that emphasize the importance of capturing rationale, alternatives, and consequences in architectural decisions \cite{shahin2009architectural}. However, unlike \citet{shahin2009architectural}, our results empirically show how this structural detailing is perceived by users with a similar profile as P13, a software engineer with 1-2 years of working experience, as increasing documentation effort and cognitive demand.

Temporal constraints emerged as a decisive factor influencing template selection, highlighted by Finding 3. Participants P15 and P03 explicitly associated short development cycles with a preference for low-overhead templates, aligning with Nygard’s original motivation to adapt architectural documentation to fast-paced industrial environments \citep{nygard2011adr}. This helps to contextualize the reports of \citet{ahmeti2024architecture} that the introduction of using ADRs improves satisfaction and knowledge transfer.

Project scale, as a business matter, further influenced participants’ preferences, with MADR favored in large-scale contexts and Nygard preferred for smaller projects, as evidenced by finding 4. This observation resonates with \citet{zimmermann2015architectural}, who argue that architectural knowledge management approaches should be tailored to project needs. Our findings extend this view by empirically identifying which structural characteristics of ADR templates are perceived as suitable for different project scales.

\section{Threats to Validity}

\textbf{Construct validity} concerns the accuracy of the operational measures to capture the intended constructs. In this study, comprehension, usability, and ease of adoption were operationalized through quantitative metrics derived from the DESMET FA method. Although this choice enables objective comparison and hypothesis testing, it may not fully capture the nuances in user perceptions that qualitative methods such as interviews or think-aloud protocols could reveal. To mitigate this threat, we grounded our metrics in established DESMET criteria and complemented the quantitative analysis with a qualitative analysis of participant feedback, which provided additional insight into template preferences.

\textbf{Internal validity} relates to confounding factors that may influence the observed outcomes. We employed a crossover design, which improves statistical power, since each participant's performance with one template is compared directly against their own performance with the other. However this design introduces potential carryover effects such as learning or fatigue. To mitigate these threats, we introduced a one-week interval between the two tasks to minimize the fatigue and learning effects. We also systematically counterbalanced the treatment order across participants. Regarding the preliminary selection phase, a potential threat is the subjective nature of the expert-based feature analysis (DESMET FA), as the evaluation was conducted by the authors rather than independent external practitioners. To mitigate this bias, the evaluators drew on their extensive industrial experience and strictly adhered to the standardized DESMET evaluation criteria to reach consensus.

\textbf{Conclusion validity} concerns the statistical analysis of results and the composition of subjects. We carefully selected statistical tests based on the characteristics of the paired samples, verifying normality assumptions and applying non-parametric tests when required. The use of a within-subjects design further increased statistical power, given the limited sample size. To address concerns regarding the sample size, we calculated non-parametric effect sizes to assess the practical significance of the results.

\textbf{External validity} refers to the generalizability of the results beyond the experimental setting. Our participants were undergraduate software engineering students enrolled in a Software Architecture course, which may limit direct generalization to experienced industrial practitioners. However, the selected participants represent junior software engineers with foundational architectural training. Prior studies have also shown that students can serve as reasonable proxies in controlled software engineering experiments when tasks reflect realistic activities~\cite{falessi2018empirical}.

\section{Conclusion}
This study evaluated ADR templates to provide empirical evidence for their usability in software projects. Using the DESMET FA framework, we narrowed a broad evaluation of five industry templates down to a controlled experiment comparing Nygard and MADR with 33 participants.

Quantitative results showed a statistically significant difference ($p = 0.002$) on the DESMET FA's overall score of the models. While both templates capture the essence of architectural choices, MADR’s granularity better supports rationale depth, whereas Nygard’s was perceived as more lightweight and easier to apply under time constraints. Our qualitative analysis suggests that template choice is context-dependent, shaped by trade-offs between rationale depth, time constraints, and project criticality. 

From a research perspective, this work contributes to the field of architectural decision documentation by providing a replicable, empirically grounded comparison of widely adopted ADR templates, complementing previous studies that focused on conceptual analyses or model descriptions. By combining professional evaluation with controlled experimentation through the DESMET FA method, the study offers evidence-based insights into how the structural characteristics of ADR templates influence their perceived usability and adoption. This could work as a hypothesis for future studies.

For practitioners, the findings suggest that ADR template selection should not follow a ``one-size-fits-all'' approach. Instead, teams tailored the template choice to their specific architectural context, time constraints, and documentation goals, acknowledging the inherent trade-offs between expressiveness and efficiency.

Future research will explore using MADR and Nygard as schemas for automated ADR generation by mining commit messages and pull request discussions. We also plan to evaluate how these templates perform within LLM-driven workflows for automated knowledge management.

\section{ACKNOWLEDGMENTS}
We thank all the participants in the empirical study. The present work was supported by: Coordenação de Aperfeiçoamento de Pessoal de Nível Superior – Brasil (AUXPE-CAPES-PROEX) – Financing Code 001; and Amazonas State Research Support Foundation – FAPEAM – through the PDPG-CAPES project. We also would like to thank the financial support granted by CNPq 314797/2023-8; CNPq 443934/2023-1; and CNPq 445029/2024-2. 

\section{ARTIFACT AVAILABILITY}
All instruments, raw data, and detailed results referenced in this
paper are publicly available and can be verified at \url{https://github.com/foolnando/ease-one-size-fits-all-adr.git}

\bibliographystyle{ACM-Reference-Format}
\bibliography{sample-base}


\end{document}